# Reciprocity between Moduli and Phases in Time-Dependent Wave-Functions

R. Englman*, A. Yahalom# and M. Baer*

*Department of Physics and Applied Mathematics,
Soreq NRC, Yavne 81800, Israel
#Faculty of Engineering. Tel Aviv University,
Ramat Aviv, Israel
(email:englman@vms.huji.ac.il)

Abstract.
For time (t) dependent wave functions we derive rigorous conjugate relations between analytic decompositions (in the complex t-plane) of the phases and of the log moduli. We then show that reciprocity, taking the form of Kramers-Kronig integral relations (but in the time domain), holds between observable phases and moduli in several physically important instances. These include the nearly adiabatic (slowly varying) case, a class of cyclic wave-functions, wave packets and non-cyclic states in an "expanding potential". The results exhibit the interdependence of geometric-phases and related decay probabilities. Several known quantum mechanical theories possess the reciprocity property obtained in the paper

1. Introduction.

The presence of i in the time-dependent Schrodinger equation

$$i\partial \Psi(x,t)/\partial t = H(x,t)\Psi(x,t) \tag{1}$$

(in which t is time, x denotes all particle coordinates, H(x,t) is a real Hamiltonian and h=1) makes the solution $\Psi(x,t)$ complex-valued. Written in the form

$$\log \Psi(x,t) = \log|\Psi(x,t)| + i\arg[\Psi(x,t)] \tag{2}$$

we have in the the first term the modulus $|\Psi(x,t)|$ which has become associated through Born with a probability, or probability density, and has soon formed the main interpretative tool of Quantum Mechanics [1]. In the second term we recognize (or define) arg as the "phase" of the wave function which expresses its signed or complex valued nature. It is recorded that at the beginning the existence of the phase was hard to accept ([1], p. 266). In rather later works the phase attracted attention that has made its significance comparable to that of the modulus. We mention the paper of Aharonov and Bohm [2] and those by Berry [3].
  In this paper we shall investigate what, if any, interrelations exist between the moduli and phases? Are they independent quantities, or, more likely since they derive from a single equation (1), are they interconnected. We shall seek exact and general relations, as in Eqs.(4) and (5) below. Some approximate and heuristic connections between phases and moduli have been known before ([4] – [8]); we shall return to these in the Discussion section.
  As will presently emerge, a central ingredient in the analysis are the zeros of $\Psi(x,t)$ (or roots of $\Psi(x,t)=0$. There are no singularities in $\Psi$ except, possibly, at points where the Hamiltonian is singular). Within quantum mechanics the zeros have not been given much attention, but they have been studied in a mathematical context [9] and in some classical wave-phenomena ([10] and references therein). Their relevance to our study is evident since at its zeros the phase of $\Psi(x,t)$ lacks definition.
It is purposeful to state the context of this paper and its aim in a setting more general than $\Psi(x,t)$, which is just a particular, the coordinate representation of the evolving state. (We shall return to this representation in section 3D). More generally we write

$$\Psi(x,t) = \Sigma_i \phi_i(t) |i\rangle \tag{3}$$

where $|i\rangle$ represent some time-independent orthonormal set and $\phi_i(t)$ are the corresponding amplitudes. We shall write generically $\phi(t)$ for any of the "component–amplitudes" $\phi_i(t)$ and derive from it, as immediately below in Eq. (6), a new function $\chi(t)$ that retains all the *fine-structure* in $\log\phi(t)$ as a function of t, and is free of the large scale variation in the latter. We then derive in several physically important cases, but not in all, reciprocal relations between the modulus and phase of $\chi(t)$ of the form

$$(1/\pi)P\int_{-\infty}^{\infty} dt'[\log|\chi(t')|]/(t'-t) = \pm \arg\chi(t) \tag{4}$$

and

$$-(1/\pi)P\int_{-\infty}^{\infty} dt' [\arg\chi(t')]/(t'-t) = \pm \log|\chi(t)| \tag{5}$$

where P denotes the principal part of the singular integral and the sign alternatives depend on the location of the zeros (or singularities) of $\chi(t)$. The above two equations are the central formulae of this work. When (4) and (5) apply, $\log|\chi(t)|$ and $\arg\chi(t)$ are "Hilbert transforms" [11].
In the next section we consider the analytic properties of the functions involved and obtain exact formulae similar to (4) and (5), but less simple and harder to apply. Still, a rather broad-ranged application is included. In section 3 we give conditions under which (4) and (5) are exactly or approximately valid. Noteworthy among these is the nearly adiabatic (slowly evolving) case, which is of special recent interest, particularly after Berry's works [3].

2. Analytic properties of the amplitude.

We suppose that $\log\phi(t)$ arising from Eq. (3) can be written in the form of

$$\log\phi(t) = P(t) + \log\chi(t) \tag{6}$$

where P(t) is a polynomial and $\log\chi(t)$ vanishes for large $|t|$ suitably fast [11]. We further write

$$\log\chi(t) = \log\chi_+(t) + \log\chi_-(t) \tag{7}$$

where $\log\chi_+(t)$ and $\log\chi_+(t)$ are respectively analytic in the upper and lower half of the complex t-plane (not including the real t-axis, where there may be singularities whose square is integrable, e.g. simple zeros of $\chi_\pm(t)$).
We now apply to the functions $\log\chi_\pm(t')$ Cauchy's theorem with a contour C that consists of an infinite semicircle in the upper/lower half of the complex t'-plane traversed clockwise/anti-clockwise and a line along the real t'-axis from $-\infty$ to $\infty$ in which the point t'=t is avoided with a small semicircle. We obtain

$$\int_C dt' \log\chi_\pm(t')/(t'-t) = \pm 2\pi i \log\chi_\pm(t') \text{ or zero} \tag{8}$$

depending on whether the small semicircle is outside or inside the half-plane of analyticity and the sign ± is taken to be consistently throughout. Writing the logarithms as

$$\log\chi_\pm(t) = \log|\chi_\pm(t)| + i \arg\chi_\pm(t) \tag{9}$$

and separating real and imaginary parts of the functions in (8) we derive the following relations between the amplitude moduli and phases in the wavefunction:

$$(1/\pi)P\int_{-\infty}^{\infty} dt'[\log|\chi_+(t')| - \log|\chi_-(t')|]/(t'-t) = \arg\chi_+(t) + \arg\chi_-(t)$$

$$= \arg\chi(t) \qquad (10)$$

and

$$(1/\pi)P\int_{-\infty}^{\infty} dt' [\arg\chi_-(t') - \arg\chi_+(t')]/(t'-t) = \log|\chi_+(t)| + \log|\chi_-(t)|$$

$$= \log|\chi(t)| \qquad (11)$$

2(A). Expanding waves.

As a first application we turn to the expanding potential problem [12-14], where we shall work from the amplitude-modulus to the phase. The time dependent potential is of the form

$$V(x,t) = \zeta^{-2}(t) V(x/\zeta(t)) \qquad (12)$$

Here $\zeta^2(t) = t^2 + c$, which differs from the more general case considered in [12-14], by putting their time scale factor a=1 and by making the potential real and time-inversion invariant. Then c is real and, in [12-13], b=0. The Hamiltonian is singular at $t = \pm i\sqrt{c}$, away from the real axis

As shown in [12], the generic form of the solution of the time dependent Schrodinger equation is the same for a wide range of potentials. We shall consider the ground state for a harmonic potential $V(x) = 1/2\, m\omega_0^2$.
The log (amplitude-modulus) of the ground state wave function (in the coordinate representation) is according to [12]

$$\log|\chi(x,t)| = -(1/4)\log(t^2+c) - \tfrac{1}{2}[m\omega x^2/(t^2+c)] \qquad (13)$$

where $\omega^2 = \omega_0^2 + c$.

The polynomial P(t) of (6) is evidently absent from (13). Processing the expression in (13) as in (6), we can be arbitrarily split it up into parts that are analytic above and below the real t-axis. Thus, let us suppose that a fraction $f_1$ of the first term and a fraction $f_2$ of the second term is analytic in the upper half and, correspondingly, fractions $(1-f_1)$ and $(1-f_2)$ are analytic in the lower half. Then substitution in the left hand side of (10) and use of an integration formula [16] give immediately

$$\arg\chi(x,t) = -(1-2f_1)(1/4)\arctan(t/\sqrt{c}) + (1-2f_2)[m\omega x^2/(t^2+c)](t/4\sqrt{c}) \qquad (14)$$

This establishes the functional form of the phase for real (physical) times. The phase of the solution given in [12,13] has indeed this functional form. The fractions $f_1$ and $f_2$ cannot be determined from our equations (9) and (10). However, by comparing with the wave functions in [12,13], we get for them the following values.

$$1-2f_1 = 2\omega/\sqrt{c}, \quad 1-2f_2 = 4\sqrt{c}/\omega \qquad (15)$$

In the excited states for the same potential, the log modulus contains higher order terms in x ($x^3$, $x^4$, etc.) with coefficients that depend on time. Each term can again be decomposed (arbitrarily) into parts analytic in the t-half-planes, but from elementary inspection of the solutions in [12,13] it turns out that every term except the lowest (shown in (12)) splits up equally (i.e. the f's are just ½) and there is no contribution to the phases from these terms. Potentials other than the harmonic can be treated in essentially identical ways.

3. Applicability of (4) and (5).

We shall now concentrate on several cases where relations (10) and (11) simplify. The most favorable case is where $\log\chi(t)$ is analytic in one half-plane say, in the upper half, so that $\log\chi_-(t)=0$. Then one obtains reciprocal relations between observable amplitude moduli and phases as in (4) and (5). Since for a potential that is free from singularities in the finite portion of the complex t-plane, solutions of the Schrodinger equation cannot be singular, singularities of $\log\chi(t)$ arise from zeros of $\chi(t)$. We turn now to the location of these zeros.

3A. The near-adiabatic limit,

We wish to prove that as the adiabatic limit is approached the zeros of the component amplitude for the "time dependent ground state " (TDGS, presently explained) are such that for an overwhelming number of zeros $t_r$, Im $t_r > 0$ and for a fewer number of other zeros $|\text{Im } t_s| << 1/\Delta E << 2\pi/\omega$, where $\Delta E$ is the characteristic spacing of the eigen–energies of the Hamiltonian , and $2\pi/\omega$ is the time scale (e.g., period) for the temporal variation of the Hamiltonian.

TDGS is that solution of the Schrodinger equation (1) that is initially in the ground state of H(x,0), the Hamiltonian at t=0. It is known that in the extreme adiabatic (infinitesimally slow) limit a system free of degeneracies stays in the ground state. We shall work in the nearly adiabatic limit, where the above is approximately, but not precisely true.

Expanding $\Psi(x,t)$ in the eigenstates $|n>$ of H(x,0), we have

$$\Psi(x,t)=\Sigma_n C_n(t) |n> \qquad (16)$$

and we assume (for simplicity' sake ) that the expansion can be halted after a finite number (say, N+1) of terms, or that the coefficients decrease in a sufficiently fast manner (which will not be discussed here). Expressing the matrix of the Hamiltonian H as $Gh_{nm}(t)$ where $h_{nm}(t)$ is of the order of unity and G positive, we obtain (with the dot denoting time-differentiation)

$$\dot{C}_n(t)=-iG\Sigma_m h_{nm}(t) C_m(t) \qquad (17)$$

The adiabatic limit is characterized by

$$|\dot{h}_{nm}(t)| << |G| \qquad (18)$$

We shall find that in the TDGS [i.e. $\Psi_g(x,t)$], the coefficient $C_g(t)$ of |g> has the form

$$C_g(t) = A_{gg}(t) \exp(-iG\varphi_g) + \Sigma'_m A_{gm}(t)\exp(-iG\varphi_m) \qquad (19)$$

Here $\varphi_m=\varphi_m(t)$ are time-integrals of the eigenvalues $e_m(t)$ of the matrix $h_{nm}(t)$,

$$\varphi_m(t) = \int_0^t e_m(t')dt' \qquad (20)$$

In the sum the value m=g is excluded and (as will soon be apparent) $A_{gm}/A_{gg}$ is small of the order of

$$|\dot{h}_{nm}(t)|/G \qquad (21)$$

To find the roots of $C_g(t) = 0$ we divide (19) by the first term shown and transfer the unity to the left hand side to obtain an equation of the form

$$1= c_1(t)\exp(-iG\delta e_1 t) +c_2(t)\exp(-iG\delta e_2 t) +\ldots \text{ (to N terms)} \qquad (22)$$

where $\delta e_1 t$, etc. represent the differences $\varphi_m-\varphi_g$ and are necessarily positive and increasing with t, for non-crossing eigenvalues of $h_{nm}(t)$. (They are written in the form shown to make clear their monotonically increasing character and are exact, by the mean value theorem, with $\delta e_1$, etc being some positive function of t.) $c_1(t)$, etc. are small near the adiabatic limit, where G is large. It is clear that (22) has solutions only at points where Im t> 0. That the number of (complex) roots of (22) is very large in the adiabatic limit , even if (22) consists of only a few number of terms, can be seen upon writing exp(-it)=z and regarding (22) as a polynomial equation in z. Then the number of solutions increases with G. Moreover, these solutions can be expected to recur periodically provided the $\delta e$'s approach to being commensurate.

It remains to investigate the zeros of $C_g(t)$ arising from having divided out by $A_{gg}(t) \exp(-iG\varphi_g))$. The position and number of these zeros depend only weakly on G, but depends markedly on the form that the time dependent Hamiltonian H(x,t) has. It can be shown that (again due to the smallness of $c_1$, $c_2$, …) these zeros are near the real axis. If the Hamiltonian can be represented by a small number of sinusoidal terms, then the number of fundamental roots will be small. However, in the t-plane these will recur with a period characteristic of the periodicity of the Hamiltonian . These are relatively long periods compared to the recurrence period of the roots of the previous kind, which is characteristically shorter by a factor of

$$|\dot{h}_{nm}(t)|/G.$$

This establishes our assertion that the former roots are overwhelmingly more numerous in number than those of the latter kind.

Before embarking on a formal proof , let us illustrate the theorem with respect to a representative, though specific example. We consider the time development of a doublet subject to a Schrodinger Equation whose Hamiltonian in the doublet representation is [16]

$$H(t) = G/2 \begin{pmatrix} -\cos\omega t & \sin\omega t \\ \sin\omega t & \cos\omega t \end{pmatrix} \quad (23)$$

Here $\omega$ is the angular frequency of the system, imposed (say) by an external disturbance. The eigenvalues of (23) are $G/2$ and $-G/2$. If $G>0$, then in the ground state the amplitude of $|g\rangle$ is

$$C_g = \cos(Kt)\cos(\omega t/2) + (\omega/2K)\sin(Kt)\sin(\omega t/2)$$
$$+i(G/2K)\sin(Kt)\cos(\omega t/2) \quad (24)$$

with
$$K = 0.5\sqrt{(G^2+\omega^2)} \quad (25)$$
$$\approx G/2, \text{ since } G/\omega >> 1$$

Thus the amplitude in (24) becomes

$$C_{gg}(t) \approx \exp(iKt)\cos(\omega t/2) + (\omega/2K)\sin(Kt)\sin(\omega t/2)$$
$$\approx e^{iGt/2}[\cos(\omega t/2) - i((\omega/2G)\sin(\omega t/2))] + ie^{-iGt/2}((\omega/2G)\sin(\omega t/2)) \quad (26)$$

This is precisely of the form (19), with the second term being lower than the first by the small factor shown in (21). Equating (26) to zero and dividing by the first term, we recover the form in (22), whose right hand side consists now of just one term. For an integer value of $G/\omega = M$ (say) which is large and $\exp(-i\omega t) = Z$, the resulting equation in Z has about M roots with $|Z|>1$ (or, what is the same, Im $t > 0$). As noted above, further roots of $C_{gg}(t)$ will arise from the neighbourhood of $\cos(\omega t/2)=0$, or $Z=-1$. The upper state of the doublet states has the opposite properties, namely about M roots with Im $t < 0$. We have treated this case in a previous work [17].

A formal derivation of the location of the zeros of $C_g(t)$ for a general adiabatic Hamiltonian can be given, following proofs of the adiabatic principle (e.g. [18-20]). The last source [20] derives an evolution operator U, which, with some slight notational change, can be shown to have the form

$$U(t) = A(t) \Phi(t) W(t) \quad (27)$$

[Eq. XVII.86 in the source]. Here $A(t)$ is a unitary transformation [Eq. XVII.70] that "takes any set of basis vectors of $H(x,0)$ into a set of basis vectors of $H(x,t)$ in a continuous manner" and is independent of G. In the previous worked example its components are of the form $\cos(\omega t/2)$ and $\sin(\omega t/2)$ [20]. The next factor in (27) is diagonal [XVII.68] and consists of terms of the form

$$\Phi(t) = \exp(-iG\varphi_m)\delta_{nm} \quad (28)$$

Finally, the unitary transformation $W(t)$ was shown to have a near-diagonal form [Eq. XVII.97]

$$\dot{W}(t) = \delta_{nm} + (I\,h(t)\,I/G)\,\delta W_{nm} \qquad (29)$$

The gg –component of the evolution matrix U is just $C_g$ and is, upon collecting the foregoing,

$$C_g(t) = \Sigma_m\, A_{gm}(t)\, \exp(-iG\varphi_m)[\,\delta_{mg} + (I\,\dot{h}(t)\,I/G)\,\delta W_{mg}\,] \qquad (30)$$

This can be rewritten as

$$C_g(t) = A_{gg}(t)\,[1 + (I\,\dot{h}(t)\,I/G) I\delta W_{gg}]\,\exp(-iG\varphi_g) \\ + (\dot{h}(t)\,I/G)\,\Sigma'_m\, A_{gm}(t)\,\exp(-iG\varphi_m)\delta W_{mg} \qquad (31)$$

with the summation excluding g. This is again of the form of (19), establishing the generality of the location of the eigenvalues for the nearly adiabatic case.

3B. Cyclic wave-functions.

This is a particularly interesting case, for two reasons. First, time-periodic potentials such that arise from external periodic forces, frequently give rise to cyclically varying states. (According to a recent reference "The universal existence of the cyclic evolution is guaranteed for any quantum system" [21].) The second reason is that the Fourier expansion of the cyclic state spares us the consideration of the convergence of the infinite-range integrals in (4) and (5); instead, we need to consider the convergence of the (discrete) coefficients of the expansion.

In this section we show that in a broad class of cyclic functions one half of the complex t-plane is either free of amplitude-zeros, or has zeros whose contributions can be approximately neglected. As already noted above, in such cases, the reciprocal relations connect observable phases and moduli (exactly or approximately). The essential step is that a function $\phi(t)$ cyclic in time with period $2\pi$ can be written as a sine-cosine series. We assume that the series terminates at the N'th trigonometric function, with N finite. We can write the series as a polynomial in z, where $z = \exp(it)$, in the form

$$\phi(t) = \sum_{m=0}^{2N} c_m\, z^{m-N} \qquad (32)$$

$$= z^{-N} c_0 \chi(t) = z^{-N} c_0 \sum_{m=0}^{2N} (c_m/c_0)\, z^m \qquad (33)$$

If $\phi(t)$ is a wave function amplitude arising from a Hamiltonian that is time-inversion-invariant then we can choose $\phi(-t) = \phi^*(t)$ for real t, where the star denotes the complex conjugate. Then the coefficients $c_m$ are all real. Next, factorize in products as

$$\chi(t) = \prod_{k=0}^{2N} (1 - z/z_k) \qquad (34)$$

where $z_k$ are the (complex) zeros of $\chi(t)$ or $\phi(t)$, $2N+1$ in number. Then the decomposition shown in (7), namely $\log\chi(t)= \log\chi_+(t)+ \log\chi_-(t)$, will be achieved with

$$\log\chi_+(t)= \sum_{k=0}^{R} \log(1-z/z_{k+}) \qquad |z_{k+}|\geq 1 \qquad (35)$$

$$\log\chi_-(t)= \sum_{k=R+1}^{2N} \log(1-z/z_{k-}) \qquad |z_{k-}|< 1 \qquad (36)$$

provided that $R+1$ of the roots are on or outside the unit circle in the z-plane and $2N-R$ roots are inside the unit circle. The results in (10) and (11) for the phases and amplitudes can now be applied directly. But it is more enlightening to obtain the coefficients in the complex Fourier-series for the phases and amplitudes. This is easily done for (35), since for each term in the sum

$$|z/z_{k+}| = |\exp(it)/z_{k+}| < 1 \qquad (37)$$

and the series expansion of each logarithm converges. (When, in (37) equality reigns, which is the case when the roots are upon the unit circle, the convergence is "in the mean" [22]) Then the n'th Fourier coefficient is simply the coefficients of the term $\exp(int)$ in the expansion, namely,
$-(1/n)(1/z_{k+})^n$.

The corresponding series-expansion of $\log\chi_-(t)$ in (36) is not legitimate, since now in every term

$$|z/z_{k-}| = |\exp(it)/z_{k-}| > 1 \qquad (38)$$

Therefore we rewrite

$$\log\chi_-(t)= -\sum_{k=R+1}^{2N} \log(-z_{k-}) + (2N-R)it + \sum_{k=R+1}^{2N} \log(1-z_{k-}/z) \qquad (39)$$

Each logarithm in the last term can now be expanded and the $(-n)$'th Fourier coefficient arising from each logarithm is $-(1/n)(z_{k-})^n$. To this must be added the $n=0$ Fourier coefficient coming from the first, t-independent term and that arising from the expansion of second term as a periodic function, namely

$$it= -2i\sum (-1)^n \sin(nt)/n \qquad (40)$$

For the Fourier coefficients of the modulus and the phase we note that, because of the time inversion invariance of the amplitude, the former is even in t and the latter is odd. Therefore the former is representable as a cosine series and the latter as a sine series. Formally:

$\log(\chi)=\log|(\chi)| + i \arg(\chi)$

$$= \Sigma_n A_n \cos(nt) + i\Sigma_n B_n \sin(nt) \quad (41)$$

When expressed in terms of the zeros of $\chi$, the sine-cos coefficients of the log–modulus and of the phase are respectively:

$$A_0 = -\sum_{k=R+1}^{2N} \log|z_{k-}| \quad (42)$$

(written in terms of the *moduli* of the roots $z_{k-}$, since the roots are either real or come in mutually complex conjugate pairs. In any case, this constant term can be absorbed in the polynomial $P(t)$ in Eq.(6).])

$$A_n = \left[\sum_{k=0}^{R} 1/(z_{k+})^n + \sum_{k=R+1}^{2N}(z_{k-})^n\right]/n \quad (43)$$

$$B_n = \left\{\sum_{k=0}^{R} 1/(z_{k+})^n - \sum_{k=R+1}^{2N}[(z_{k-})^n - 2(-1)^n]\right\}/n \quad (44)$$

Equations (42)-(44) are the central results of this section. Though somewhat complicated, they are easy to interpret, especially in the limiting cases (a) –(d), to follow. In the general case, the equations show that the Fourier coefficients are given in terms of the amplitude zeros.

(a) When there are no amplitude zeros in one of the half planes, then only one of the sums in (43) or (44) is non-zero (R is either 0 or 2N). Consequently, the Fourier coefficients of the log modulus and of the phase are the same (up to a sign) and the two quantities are logically interconnected as functions of time. The connection is identical with that exhibited in Eqs. (4) and (5).

   In the two-state problem formulated by (23), the solution (24) is cyclic provided $K/\omega$ is an integer. A "Mathematica" output of the zeros of (24) is shown in Fig.1 for $K/\omega = 8$. It is seen that none of the zeros is located in the lower half plane: 7 pairs and an odd one is in the upper half plane proper, a pair of zeros is on the real t-axis. The reciprocal integral relations in (4) and (5) are verified numerically, as seen in Fig. 2. (The equality between the Fourier coefficients $A_n$ and $B_n$ was verified independently.)

(b) It is a characteristic of the above two state problem (with general values of $K/\omega$) and of other problems of similar type that there is one or more roots at or near $z_{k+} = -1$ ($t = \pm\pi$; the generality of the occurrece of these roots goes back to a classic paper on conical intersection [23].). By inspection of the second sum in (44) we find that, if all the roots located in the upper half plane are of this type, then $A_n = B_n$ up to small quantities of the order of $(z_{k+}+1)$. Then again Eqs. (4) and (5) can be employed..

(c) If either $|z_{k+}| \gg 1$ or $|z_{k-}| \ll 1$, it is clear from (43) and (44) that the contribution of such roots is small. This circumstance is important for the following reason: Suppose that the model is changed slightly by adding to the potential a small term, e.g., adding $\varepsilon\cos 2\omega t$ to a diagonal matrix element in (23), where $\varepsilon$ is small. (In Ref. [24] terms of this type were used to describe the nonlinear part of a Jahn-Teller effect.) Necessarily, this term will introduce new zeros in the amplitude. It can be shown that this addition will add new roots of the order $|z_{k+}| \approx 1/\varepsilon$ or $|z_{k-}| \approx \varepsilon$. The effects of these are asymptotically negligible. In other words, the formulae (43) and (44) are stable with respect to small variations in the model. (A similar result is known as Rouche's theorem about the stability of the number of zeros in a finite domain, [9] Section 3.42)

3D. Wave packets

A time varying wave function is also obtained with a time-independent Hamiltonian by placing the system initially into a superposition of energy eigenstates (|n>), or forming a wave-packet. Frequently a coordinate representation is used for the wave function which then may be written as

$$\Psi(x,t) = \Sigma_m a_m \exp(-iE_m t) \psi_m(x) \qquad (45)$$

where $\psi_m(x)$ are solutions of the time independent Schrodinger equation, with eigenenergies $E_m$ that are taken as non-degenerate and increasing with m. In this coordinate-representation, the "component-amplitudes" in the Introduction are just fancy words for $\Psi(x,t)$ at fixed x (so that the discrete state label n in the Introductory section is equivalent to the continuous variable x) and $\phi_n(t)$ is simply $\phi_x(t) \equiv \Psi(x,t)$. The results in the earlier section are applicable to the present situation. Thus, to test (4) or (5), one would look for any fixed position x in space at the moduli (or state populations) as a function of time, as with repeated state-probing set ups. In turn, by some repeated interference experiments at the same point x, one would establish the phase and then compare the results with those predicted by the equations. (Of course, the same equations can also be used to predict one quantity, provided the time history of the second is known.)

As in previous sections, the zeros of $\Psi(x,t)$ in the complex t-plane at fixed x are of interest. This appears a hopeless task, but the situation is not that bleak. Thus, let us consider a wave packet initially localized in the ground state in the sense that in (45), for some given x,

$$\sum_{m>0} |a_m \psi_m(x)|^2 < |a_0 \psi_0(x)| \qquad (46)$$

Then we expect that for such value(s) of the coordinate the t- zeros of the wave-packet will be located in upper t-half plane only. The reason for this is similar to the reasoning that led to the theorem about the location of zeros in the near-adiabatic case. (Section 3A)

Actually, empirical investigation of wave-packets appearing in the literature indicates that the expectation holds in a broader range of cases, even when the condition (46) is not satisfied. It should be mentioned that much of the wave packet work is numerical and it is not easy to theorize about it. (A review describing certain aspects of wave-packets is found in [25].)

We now present some examples of studied wave-packets for which the reciprocal relations hold (exactly or approximately), but have not been noted.

### (a) *Free-particle in one-dimension.*

The Hamiltonian consists only of the kinetic energy of the particle having mass m ([26], [5] Section 28). The (unnormalized) energy eigenstates labelled by the momentum index k are

$$\psi_k(x) = \exp(ikx) \tag{47}$$

with corrresponding energies $E_k = k^2/2m$. Initially the wave packet is centered on x=0 and has mean momentum K. For a minimum uncertainty product wave packet, the coefficients $a_k$ shown in (45) are

$$a_k = \exp[-(k-K)^2 \Delta^2] \tag{48}$$

where $\Delta$ (>0) is the root mean square width in the initial wave packet. The expanding wave packet can be written as

$$\log\Psi(x,t) = -\tfrac{1}{2}\log[\Delta + (it/2m\Delta)] - [x^2 - 4i\Delta^2 K(x - Kt/2m)]/[4\Delta^2 + 2it/m] + \text{constant} \tag{49}$$

which is clearly analytic in the lower half t-plane. We can therefore identify this function with $\log\chi(t) = \log\chi_-(t)$ in (7), and tae $\log\chi_+(t) = 0$. As a numerical test we have inserted (49) in (4) and (5) integrated numerically and found (for K=0) precise agreement. (Fig.3)

### (b) *"Frozen Gaussian Approximation".*

Semi-analytical and semi-classical wave packets suitable for calculating evolution on an excited state multi-dimensional potential energy surface were proposed in pioneering studies by Heller [32]. In this method (called the Frozen Gaussian Approximation) the last *two* factors in the summand of (45) were replaced by time dependent gaussians. The time dependence arose through having time varying average energies, momenta and positions. Specifically, each coefficient $a_m$ in (45) was followed by a function g(x,t) of the form

$$\log g(x,t) = -m\omega(x - \langle x\rangle_t)^2/2 + i\langle p\rangle_t(x - \langle x\rangle_t) + i\int_0^t (\langle p\rangle_t^2/m - \langle g|H|g\rangle_t)\,dt \tag{50}$$

where $\omega$ is an energy characteristic of the upper potential surface, the angular brackets are the average position and momenta of the classical trajectory and the Dirac bracket of the Hamiltonian H is to be evaluated for each component g separately.

For a *set* of gaussians it is rather difficult to establish the analytic behavior of (50), or of (45), in the t-plane. However, with a *single* gaussian (in one spatial dimension) and a harmonic potential surface one has classically

$$\langle x\rangle_t = x_0\cos\omega t, \tag{51}$$

$<p>_t = (m)d<x>_t/dt = -(\omega x_0 m)\sin\omega t$ (52)

$<g|H|g>_t = \omega/2$ (53)

Substituting these expressions into (50) one can see after some algebra that log g(x,t) can be identified with $\log\chi_-(t) + P(t)$ shown in section 2. Moreover, $\log\chi_+(t) = 0$. It can be verified, numerically or algebraically, that the log-modulus and phase of $\log\chi_-(t)$ obey the reciprocal relations (4) and (5).
In more realistic cases (i.e., with several gaussians) (51)- (53) do not hold. It still may be true that the analytical properties of the wave packet remain valid and so do relations (4) and (5). If so, then these can be thought of as providing numerical checks on the accuracy of approximate wave-packets.

4. Discussion

As a prelude to a discussion of our results we note here some of the relations between phases and moduli that have been known previously. The following is a list (presumably not exhaustive) of connections some of which are standard text-book material.

(a)     The Equation of Continuity.

This was first found by Schrodinger in 1926 using Eq.(1), which he called the "eigentliche Wellengleichung", [In translation "the real (sic) wave equation" [4]]. In the form

$2m\partial \log|\Psi(x,t)|/\partial t - 2\text{grad} \log|\Psi(x,t)|\cdot\text{grad arg}[\Psi(x,t)] - \text{divgrad arg}[\Psi(x,t)] = 0$ (54)

(where m is the particle mass), it is clearly a (differential) relation between the modulus and the amplitude. However, the above form depends on the Hamiltonian and looks completely different for, e.g, the Dirac equation.

(b) The WKB formula.

In the classical region of space, where the potential < the energy, the standard formula leads to the following path integral relation, between phase and modulus, ([5] , Section 28)

$$\text{Phase} = \pm C\int_0^{x(t)} |\Psi(x)|^{-2} dx \quad (55)$$

where C is a normalization constant. This and the following example do not arise from the time dependent Schrodinger equation; nevertheless, time enters naturally in a semi-classical interpretation.

(c) Extended systems

Based on a previous heuristic proposal [6], the phase in the polarized state of a one-dimensional solid of macroscopic length L was expressed in [7] as

$$\text{Phase} = \text{Im} \log \int_0^L \exp(2\pi i x/L) \, |\Psi(x)|^2 \, dx \tag{56}$$

It has been noted that the phase in (56) is gauge-independent [7]. Based on the heuristic conjecture (but fully justified, to our mind, in the light of our rigorous results), Resta has noted that "Within a finite system two alternative descriptions [in terms of the squared modulus of the wave function, or in terms of its phase] are equivalent" [28].

(d)    Loss of phase in a quantum measurement.

In a self-consistent analysis of the interaction between an observed system and the apparatus (or environment), Stern et al [8] proposed both a phase-modulus relationship [Eq. (3.10)] and a deep lying interpretation. According to the latter, the decay of correlation between states in a superposition can be seen, equivalently, either as the effect of the environment upon the system or of the back reaction of the system on its environment. In the (microscopic) systems considered in this paper, there is only a *change* of correlation not a *decay*. Still it seems legitimate to speculate that the dual representation of the change that we have found (namely, through the phase or through the modulus) is perhaps an expression of the reciprocal effect of the coupling between the system (represented by its states) and its environment (acting through the potential).

The relations in (4) and (5) are formally identical to the well-known Kramers-Kronig relations (between real and imaginary parts of response functions), which operate in the frequency plane. The origin of the latter is in the principle of causality, leading to a real-time asymmetry (due to the imposition of initial conditions). In contrast, the relations found in this work arise from the structure of the Schrodinger equation which contains an imaginary-time asymmetry.

At this stage the following appear to be the theoretical significance of the reciprocal relations :
(i) They show that changes (of a nontrivial type) in the phase imply necessarily a change in the occupation number of the state components and vice versa.
(ii) One can define a phase that is given as an integral over the log of the amplitude modulus and therefore is an observable and is gauge-invariant.
(iii) Experimentally, phases can be obtained by measurements of occupation probabilities of states using (4). (We have verified this for the case treated in [29].)
(iv) Conversely, the implication of (5) is that a geometrical phase appearing on the left-hand-side entails a corresponding geometrical probability, as shown on the right-hand-side. Geometrical transition probabilities have been predicted in [30] and experimentally tested in [31].

Future theoretical work will focus on a systematic description of the location of zeros, of which the present work is a beginning.

References.

Captions.

Fig. 1

Zeros of the ground state wave function component $C_g(t)$, in (24) for $K/\omega=8$ (a periodic case). The roots are shown, in the form (Re t) + (Im t) I. All zeros are in the upper half of the complex t-plane, including the real axis.

| 3.141 59 | 3.141 59 |
|---|---|
| −2.675 44 + 0.131 7736i | 2.675 44 + 0.131 736i |
| −2.274 94 + 0.170 594i | 2.274 94 + 0.170 594i |
| −1.879 39 + 0.198 225i | 1.879 39 + 0.198 225i |
| −1.486 12 + 0.222 569i | 1.486 12 + 0.222 569i |
| −1.095 15 + 0.247 513i | 1.095 15 + 0.247 513i |
| −0.708 66 + 0.276 858i | 0.708 66 + 0.276 858i |
| −0.336 24 + 0.314 448i | 0.336 24 + 0.314 448i |
| 0.340 873 | |

Fig. 2

Numerical test of the reciprocal relations in Eqs. (4) and (5) for $C_g$ in Eq. (24). The values computed directly from Eq. (24) are plotted upwards and the values from the integral downwards (by broken lines). $K/\omega=8$. They are clearly identical.
2(a) Log $|C_g(t)|$ against (t /period). The modulus is an even function of t.
2(b) arg $C_g(t)$ against (t/period). The phase is odd in t.

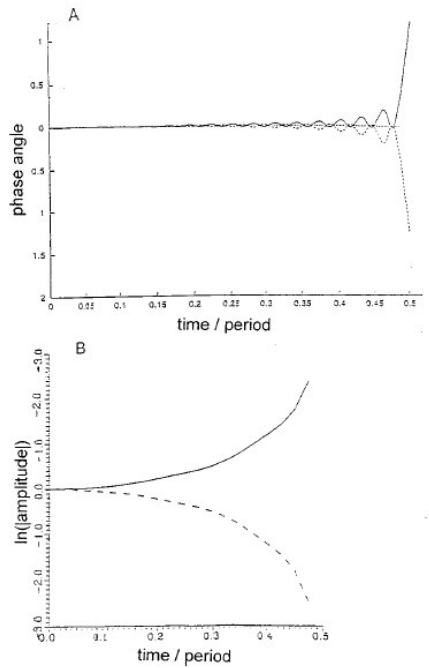

Fig. 3
Time development of a free electron wave packet, Log Ψ(x,t) in Eq. (49), for a fixed x (=1) and as function of time. The directly computed values and those obtained from the reciprocal relations in Eq. (4)- (5) are indistinguishable.
3(a) Re LogΨ(x,t),
3(b) Im LogΨ(x,t).

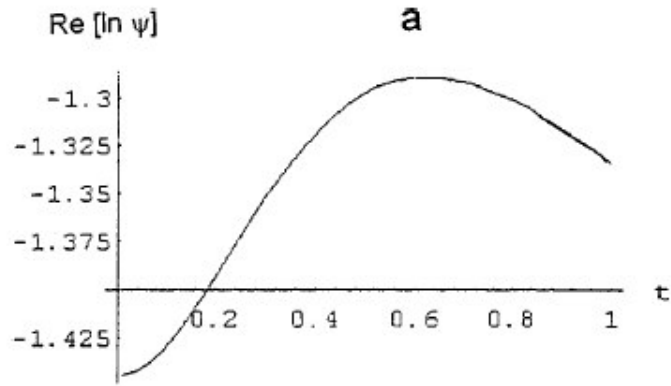

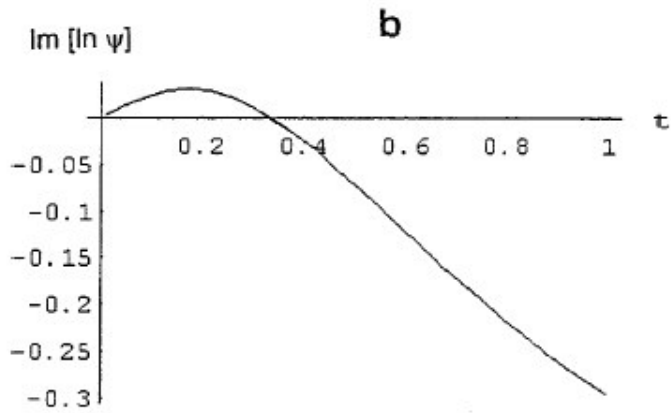